# Open Source Android Vulnerability Detection Tools: A Survey

Keyur Kulkarni
Student, EECS Dept.
The University of Toledo
Toledo, OH-43606
keyur.kulkarni@rockets.utoledo.edu

Ahmad Y Javaid
Assistant Professor, EECS Dept.
The University of Toledo
Toledo, OH-43606
ahmad.javaid@utoledo.edu

*Abstract*—Since last decade, smartphones have become an integral part of everyone's life. Having the ability to handle many useful and attractive applications, smartphones sport flawless functionality and small sizes leading to their exponential growth. Additionally, due to the huge user base and a wide range of functionalities, these mobile platforms have become a popular source of information to the public through several Apps provided by the DHS Citizen Application Directory. Such wide audience to this platform is also making it a huge target for cyber- attacks. While Android, the most popular open source mobile platform, has its base set of permissions to protect the device and resources, it does not provide a security framework to defend against any attack. This paper surveys threat, vulnerability and security analysis tools, which are open source in nature, for the Android platform and systemizes the knowledge of Android security mechanisms. Additionally, a comparison of three popular tools is presented.

## I. Introduction

In the modern world, smartphones have become a necessity for everyone. Late 1990s saw a growth in use of PDAs (personal digital assistant) and it did not take much time to transform them into mobile devices, popularly known as smartphones. With abilities of PCs, these devices offer numerous appealing features such as connectivity (GPRS, LTE, GSM, Bluetooth, NFC), multi-tasking, storage, fingerprint sensor, wireless charging, and beautiful graphical user interface. It is estimated that subscription of smartphones to increase to 2.5 billion in 2017. Recent years has seen a huge development in mobile device operating systems such as Android, iOS, BlackBerry and Windows Mobile. It is well know that Android - an open access OS developed by Google - has seen huge demand in recent years and has a market share of roughly 87.5%. Android being an open-source technology, it is not only limited to smartphones, but it has extended its reach to TV, car and automation systems. These devices are now the ideal target for attackers because of the huge user base and lack of presence of several security features. The number of application developers increases day by day, however, many of them lack expertise for security implementations. Many developers focus on consumers demand rather than security of the application.

Mobile security is an emerging area of research. The past 2-3 years have seen many research works of mobile security not only related to Android but also iOS. The openness and extensibility of Android have made it a popular platform for mobile devices and a strong candidate to drive the Internet-of-Things. After some research, researchers have presented some good security frameworks. These frameworks have either static or dynamic approaches. The main advantage of these approaches is that it does not require any changes to the Android code, and the hardening tools can be easily deployed as standard applications on the device.

According to Kaspersky Security Bulletin 2016 [1], most hazardous and well-known mobile Trojans were advertising Trojans. These Trojans were able to use superuser rights on infected device. Kaspersky Lab, in October 2015 detected a Trojan, which looked like a music player application but used for stealing users credentials. In 2016, ransom Trojans were also popularly used by cybercriminals. These Trojans like Trojan-Ransom.AndroidOS.Fusob, were distributed in countries like United States, Germany and United Kingdom and usually demanded ransom between 100$ or 200$, to unblock device. In China another ransom family, Trojan-Ransom.AndroidOS.Congur, blocks infected device in a unique way. Once started, it will ask for Device Administrators rights and then changes pin code or sets one if not set before. Cybercriminals will ask user to contact them through QQ messenger to get new pin code. New category of Trojan seem to be immersed in 2016, which will attack the routers controlling wireless network. This malware (also known as Switcher) aims at attacking the router's admin access using long, predefined list of login ids and passwords. If successful, the attackers will change Domain Name Servers (DNS) settings in order to re-route DNS query. Hence, the attacker will be able to monitor network traffic. Security firm Blue Coat Systems discovered a malware that targets the device without using an installed app or APK (Android application package). This malware requires to visit a particular website and device gets infected as the malware takes advantages of various vulnerabilities in OS. This particular malware only affects devices, which have Android 4.0 through Android 4.3, but that still makes up for millions of Android phones out there and is more like ransomware. Whenever someone has been a victim, that person might look to anti-virus as a solution. However, anti-virus like *Android defender* are not actually real but some fake apps designed to cause harm. These apps have convincing GUI designed to fool user. Recent news report suggests, new President of USA uses an android device and experts believe this might lead big troubles to nation as android is prone to lot of security issues [2]. DHS provides several apps to the US citizens. Since many of these apps might be storing confidential user information, it is imperative that they are not accessible to other vulnerable or insecure applications as well as attackers. To ensure this, it is necessary to vet any app installed on a phone which has such confidential user information stored on it [3].

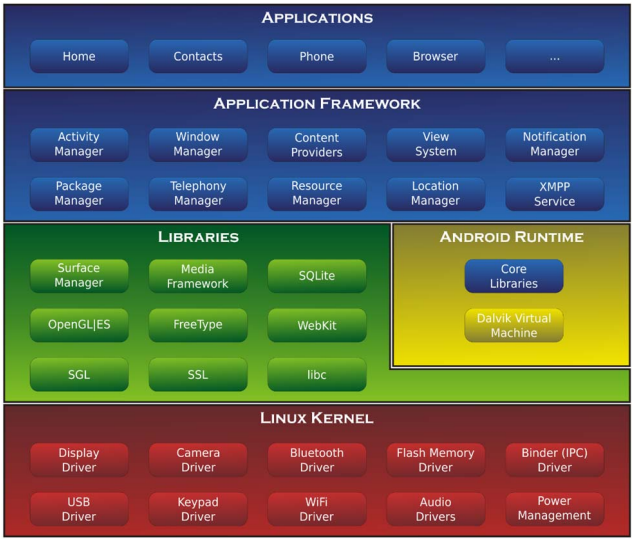

Fig. 1 Android Architecture

These problems can be addressed by performing Taint analyses on developing or developed application; it can analyze application both statically and dynamically. While static analysis imposes no run-time overhead, it is inherently imprecise. In contrast, dynamic taint-tracking, which uses *taint tags* to track whether a value contains private information has been shown to be effective at detecting personal privacy violations. For example, TaintDroid can track taints in real-time with a mere 14% performance overhead. As a result, TaintDroid gains great popularity among mobile device users [4].

The rest of paper is organized as follows. Section II gives overview of android architecture and security. Section III describes in brief various tools available at ones disposal. In Section IV comparison through experiments of some common tools is presented.

## II. ANDROID, SECURITY THREATS & FEATURES

### A. Android Overview

The overall Android system architecture is shown in the Fig. 1. Architecture has four main layers, providing services for the layer above and using services of the layer below. Android OS built on Linux kernel, allows being ported to various platforms with the help of hardware abstraction. Android native libraries in C or C++ are compiled by particular hardware architecture used by the device and lie above Linux kernel. Android consists of Dalvik Virtual Machine (DVM) and core Java libraries. Application framework layer enables control of lifecycle of an application, shares data between applications, provides a geographic location using GPS and manages notification. The uppermost layer consists of application and widget.

Android has a cohesive approach to application development for mobile devices allowing the developed application to run on different devices driven by Android. Android applications can be considered the main source of the attack. Even precisely written application can be a source of malicious activity such as a leak of sensitive or private data, in form of messages, contacts, location, etc.

TABLE I. STRIDE THREAT MODEL

| Threat | Definition | Property |
|---|---|---|
| **S**poofing | Impersonating user | Authentication |
| **T**ampering | Modifying data or code | Integrity |
| **R**epudiation | Perform action which is not attributable | Non-Repudiation |
| **I**nformation Disclosure | Expose information to unauthorized person | Confidentiality |
| **D**enial of Service | Deny access to valid users | Availability |
| **E**levation of Privilege | Gain access without proper authority | Authorization |

### B. Security Threats

The growing popularity of smartphones, charms attention of cybercriminals who try to seize this opportunity in their hand for information and money. Latest OSes have been designed keeping in mind the security, so the easiest path for an attack is malware. Cybercriminals are now using attacks and techniques initially targeted at desktop users in the mobile channel. They are experts at social engineering and are executing targeted spear phishing attacks. Android devices are under attack by cybercriminals mostly because of [5]:

- App side-loading (allow third-party apps to be installed)
- Hacking an Android device is easy for cybercriminals
- Weak rules governing app signing

The vulnerability is a weakness in a system that aids attackers in the successful execution of threat or exploits and Risk is threat times vulnerability (potential loss/damage because of threat exploitation using vulnerability). Threat modeling allows to analyze the security of the application and it helps in understanding entry points and respective threats associated with them. STRIDE is a threat model used to identify threats and categorize them. Table I describes model for mobile along with categorized threats. The attacks on smartphones can be distinguised as [6] [7]:

1) *Malware:* Malware or malicious software, designed to target the device such as a smartphone with the intention to damage or disrupt it and can be grouped into categories like a virus, worm, Trojan, rootkits, etc. It attacks the device without users consent. Symptoms like crashing, problems connecting to network, modified or deleted files gives presence of malware.
2) *Grayware:* Grayware refers to applications that are not classified as viruses or Trojan-horse programs, but can still negatively affect the performance of the system and introduce significant security risks to the organization. Often grayware performs a variety of undesired actions such as irritating users with pop-ups, tracking user habits and unnecessarily exposing vulnerabilities to attack.
3) *Spyware:* Spyware collects user information and sends to its developer/attacker. It usually hides itself from the user and it gets downloaded from 'free downloads' and doesn't even ask for users permission to install. Software such as Adware can be categorized in spyware.

HTTP protocol is used for communication and in few cases use of C&C via Google Cloud is observed. In-built SMS functionality can also be used. Once malware gains access to device, it can harm in one or many of following ways [8]:

- wipe device
- reset lock screen PIN
- open an arbitrary URL in the phone's browser
- send an SMS message to any or all contacts
- lock or unlock the device
- steal received SMS messages
- steal contacts
- display a different ransom message
- enable or disable mobile data
- track user's GPS location

Google's Play Store is a digital platform for distribution of Android application(s). According to experts, Google Play is not a well-policed platform and is one of main source of vulnerable application. Malware is known to be located in free app or free version of commercial apps on third-party app stores. The user is lured into installing the application that has malicious contents. For example, advertisements running through an application can be used to lure users into installing malicious applications. These risks can be more aggravated because of Android's fragmentation of devices and OS. Once Google releases a version of OS, manufacture or carriers may not upgrade their OSes or may take some time for upgradation, hence a device may run on an unstable version of OS having security issues. If device runs on older version of OS, which means certain vulnerabilities will remain unpatched leaving an area for an attacker to exploit.

Android's open market policy makes it easy to access the app from alternate markets besides Play Store. These markets have a weak background check on app issuers or developers. There even exists a way to install the application without the use of app stores; it can be done with the help of "adb tool" when the device is connected to the computer. These give the user more freedom to install apps from non-market sources, but it opens up an additional entry point for malware. Android apps are relatively easier to reverse-engineer compared to native apps in a desktop environment, such as Windows and UNIX executables, because hardware independent Dalvik bytecode files retain a great deal of information of the original Java sources. Android applications lack runtime ICC (inter-component communication) control for purposes like (i). Prevent app from intercepting intent broadcast and possibly stopping its propagation afterward, (ii) to prevent the app from accessing any open interfaces of another app and to isolate apps and (iii) prevent them from communicating via ICC and other shared channels. JNI (Java Native Interface) can be used to invoke an app, this cause bug issues or memory corruption at low-level languages. There can also be issues with pre-installed apps and manufacturer's customization.

A software that will download and display ads when the user is online is known as an Adware. They collect user information without the knowledge of the user and share the collected information with advertising agencies. SDKs are offered by Ad networks which developer can embed into their application; this application will make use of IMSI and IMEI of the device to identify it to the server and share vital data. Trojans such as info stealers gather confidential information by monitoring internet activities for banking and social media activities. Spy phone apps can be used to spy on phone's owner. They track the phone's location, monitor incoming and outgoing calls, text messages, email and track the victim's browsing. SMS Trojan can send text messages from compromised phone to other premium SMS numbers for which phone's owner has to pay.

Android allows installation of app from third-party source, which can be potentially harmful to the device, its user or user data. Potentially Harmful Applications (PHAs) attacks can range from data collection for targeted advertisement to potentially driven attacks for users' harm. Information Extraction is a possible attack. Applications can have access to IMEI number through API calls if the application has proper rights. More or less you can get IMEI numbers are available in black markets. Many vendors provide features to their consumers like free calling and SMS, but some few services are premium. An attacker can easily make use of these services, make calls or send SMS to such premium numbers and the consumer or user has to pay for it. An example of an app is *Fakeplayer* [9]. Search Engine Optimization another form attack, which uses fraudulent clicks on the target website. The attacker can use users' device to rise a particular website in search results. One of loop hole in Android security is that Android application can be seen downloading payloads or executable comprising of native codes or libraries for code execution. It can be possible for an attacker to make the malicious code available instead of original native code. Users can be seen customizing their devices to gain certain access to resources by rooting the device. Both genuine users and malware attackers carry out Root exploits. Malwares are granted privileges with the help of exploits such as *Exploid* and *Zimperlich* [9].

*C. Security Features*

The Android application makes use of cutting-edge software and hardware along with local and served data to bring invention and significance to the user. The platform must offer an application with a secure environment, which guarantees the security of users' device, data, application, and network. The security system should not only reduce the probability of attack but also limit the impact of the successful attack. Flexibility and Adaptability features of Android OS makes it unique as compared to its competitors like Apple iOS. Android's customizable feature makes it more insecure. Device vendors can make changes to the existing version of OS to suit their hardware specifications. Although being a great feature, it can be a source of malicious activity. Security features of Android include [5]:

- Security at OS levels through Linux Kernel
- Application Sandboxing
- Inter-process communication security
- Application signing
- User application permissions

The Linux system over the years itself has been researched for vulnerability, attacked upon and fixed by developers. Linux is used in many security sensitive applications being one of most stable and secure kernel. In Android, Linux is used as a base for mobile computing environment and provides features such as permissions model, process isolation. Android does not allow one application to use resources of another application. This is done by giving each application a unique user id (UID), and application is run as a separate process. This sets up kernel-level application sandbox. The key components of Android Security Program include [5]:

- Design Review: The engineering and security resources review each feature, in order to test the outcome of previous activity or lifecycle.
- Penetration Testing and Code Review: It has to be carried out during the development of the app to identify weakness and vulnerability.
- Open source and community review: Android being open source any third party can review the app and give a response back to Google on the forum.

- Incident Response: After the report of a certain bug, Android has a response that enables rapid mitigation of threats to ensure the security of users.
- Monthly security updates: Regular and secure update to ensure the security of applications.

Taking inspiration from tools mentioned in section III, Google developed its security mechanisms for application security known as *Bouncer* [10]. Before the application is published on the Google Play, the application has to undergo a review process to verify that, app complies with policies of Google Play. Google provides security services that are automatically included as part of GMS (Google Mobile Services). The services include cloud-based services and on-device services. These services showed a record of protecting around 1 billion mobile devices in 2015. On-devices services include Verify Apps (Protection from PHA), Safety Net (Protection from network and application based threats), Android Device Manager (Protection from lost and stolen devices), Safebrowsing (protect from unsafe websites), Smart Lock (Improve user authentication and physical protection). Cloud-based services involve Static & Dynamic Analysis, Third Party Reports, Signatures, and so on along with the use of various algorithms to see patterns and make connections that a human could not.

## III. APPLICATION SECURITY ANALYSIS

Android application has two major sources: pre-installed applications and user installed applications. Android phones come with a bunch of pre-installed applications such as phone, calendar, email, contacts, etc. which provide a base for other applications. The user can install third-party applications because of open source development. Android has three types of applications: Native, Hybrid, and The Web. Native apps are built for particular devices, and coded in languages like JAVA. They can take full advantage of device features like camera, contacts, GPS. Web apps are not a real application but are websites that look like native apps. They run in browsers and written in HTML. Hybrid apps are part native and web apps. They can take advantage of device features like native apps and rely on HTML being rendered in the browser. Android application distributed from any source is in form of APK file. APK archive typically contents following directories and files:

- META-INF directory
- lib (directory having compiled code)
- res (contains resources)
- assets (containing application assets)
- AndroidManifest.xml (name, version, etc. of app)
- classes.dex (classes compiled in des file format)
- resources.arsc (precompiled resources)

Android is a component–based system and makes use of ICC, in which component can interact with one another with help of *intent*. Four kinds of components a developer can define: UI of app is implemented by *Activity* component, broadcasting message component to component is handled by *Broadcast Receiver*, background tasks performed by *services* and *providers* define a database like storage. In android app the components are invoked through callback method ('main' method does not exist in android app). Target of ICC is explicitly defined in intent or is implicit and decided at runtime.

### A. Static Analysis

Detecting and stopping threats discussed in Section II is a difficult task. Even though one follows the specific guidelines for designing an application, there is no guarantee that it will be safe. Static Analysis can be used as a tool to detect threats and rectify them. In static analysis, the application's code or source code is analyzed for any malicious activity without requiring its execution. Static Analysis involves extracting information from packaged apk's viz. source code needs to be extracted from apk. DroidMat, dex2jar, Procyon, etc. are the tools used for developing a proper tool for analysis. Recent years has shown advancement in static analysis, but these are not precise enough to use in practice. Android application run within the Android framework, this creates imposing of complex lifecycle on the apps, invoking call-back methods pre-defined or user-defined different times during execution of app. The tool must predict apps control flow and hence requires precise lifecycle modeling. Imprecise lifecycle modeling will cause the tool to lose or overshadow important data flows. Static Analysis tools have several approaches depending upon precision, runtime, scope, and focus.

CHEX [11] a tool to detect component hijacking vulnerabilities in the app, is considered as most sophisticated. It tracks taints between externally accessible interfaces and sensitive source or sink. Communication-based vulnerabilities can be detected using ComDroid [9, 12]. The tool analyzes Dalvik executable files, performs flow-sensitive intraprocedural analysis and examines app for permissions. SCandroid [9] a tool for generating automated security certification for the app using data flow analysis. Androwarn a tool for detection and warning user of malicious activities such as telephony service abuse, geolocation information leak, Denial of Service, etc. QARK (Quick Android Review Kit) is a tool having the capacity to find a common vulnerability in Android app. It is different from most of the traditional tools; it will point at vulnerabilities along with its feature of adb commands. QARK attempts to find, Weak or improper cryptography use, Improper x.509 certificate validation, Activities which may leak data, Sending of insecure Broadcast Intents, Insecurely created Pending Intents and many more security vulnerabilities.

FlowDroid [12] the first fully context, object and flow sensitive analysis. It analyzes apps bytecode and configuration files to find leaks. It is used to identify data loss caused by either carelessness or malicious intent. Amandroid [13] takes inspiration from FlowDroid; their model extends FlowDroid by capturing control and data dependencies among components. Flow of Amandroid includes conversion of apps Dalvik bytecode to intermediate representation, generate environment model, build inter-component data flow (IDFG), build data dependence graph (DDG) and then it applies security detection. Amandroid can be used for data leak detection, data injection detection and API misuse. DidFail (Droid Intent Data Flow Analysis for Information Leakage) combines and augments FlowDroid and Epicc (which identifies properties of intents such as its action string) to detect leaks within a set of Android apps. FlowDroid is extended by IccTA to analyze inter-component dataflow. LeakSemantic [14] is a hybrid static-dynamic analysis approach, to locate subtle network transmissions in apps. Agrigento [15], a new approach that is resilient to such obfuscations and, in fact, to any arbitrary transformation performed on the private information before it

TABLE II. OPEN SOURCE AVAILABILITY OF TOOLS [26] [20] [25]

| Method | Tool | Open Source (Publicly available) |
|---|---|---|
| Static Analysis | CHEX | No |
| | SCandroid | Yes |
| | Flowdroid | Yes |
| | Amandroid | Yes |
| | QARK | Yes |
| | ComDroid | No |
| | Agrigento | Yes |
| | HornDroid | Yes |
| | LeakSemantic | No |
| | Mamadroid | No |
| Dynamic Analysis | ARTDroid | Yes |
| | VetDroid | No |
| | DroidScope | Yes |
| | CopperDroid | No |
| Machine Learning approach | DroidScribe | No |
| | StormDroid | No |
| | RevealDroid | Yes |
| Real-time Monitoring | ARTist | No |
| | TaintDroid | Yes |
| | AppsPlayground | Yes |
| | Andromaly | No |
| Testing (GUI) | Sapienz | Yes |
| | Dynodroid | Yes |
| | Collider | No |

TABLE III. DROIDBENCH TEST RESULTS

O = TRUE POSITIVE, * = FALSE POSITIVE, X = FALSE NEGATIVE [13] [14]

| App Name | Flowdroid | Amandroid | QARK | Warnings |
|---|---|---|---|---|
| **Arrays Lists** | | | | |
| ArrayAccess1 | * | * | * | 3 |
| ArrayAccess2 | * | * | * | 3 |
| ListAccess1 | * | * | * | 3 |
| **Callbacks** | | | | |
| AnonymousClass1 | O | O | O | 3 |
| Button1 | O | O | O | 3 |
| Button2 | OOO* | OOO | O | 3 |
| LocationLeak1 | OO | OO | O | 3 |
| LocationLeak2 | OO | OO | O | 3 |
| MethodOverride1 | O | O* | O | 3 |
| **Field and Object Sensitivity** | | | | |
| FieldSensitivity3 | O | O | O | 3 |
| InheritedObjects1 | O | O | O | 3 |
| **Inter-App Communication** | | | | |
| IntentSink1 | X | O | * | 4 |
| IntentSink2 | O | O | O | 3 |
| ActivityCommunication1 | O | X | O | 3 |
| **Lifecycle** | | | | |
| BroadcastreceiverLifecycle1 | O | O | O | 3 |
| ActivityLifecycle1 | O | O | O | 3 |
| ActivityLifecycle2 | O | O | O | 3 |
| ActivityLifecycle3 | O | O | O | 3 |
| ActivityLifecycle4 | O | O | O | 3 |
| ServiceLifecycle1 | O | O | O | 2 |
| **General JAVA** | | | | |
| Loop1 | O | O | O | 3 |
| Loop2 | O | O | O | 3 |
| SourceCodeSpecific1 | O | O | O | 3 |
| StaticInitialization1 | X | O | O | 3 |
| **Miscellaneous Android-Specific** | | | | |
| PrivateDataLeak1 | O | O | O | 3 |
| PrivateDataLeak2 | O | O | O | 3 |
| DirectLeak1 | O | O | O | 3 |
| **Implicit Flows** | | | | |
| ImplicitFlow1 | XX | XX | X | 3 |
| ImplicitFlow2 | XX | XX | X | 3 |
| ImplicitFlow3 | XX | XX | X | 3 |
| ImplicitFlow4 | XX | XX | X | 3 |
| **Sum, Precision and Recall - DroidBench** | | | | |
| O, higher is better | 26 | 27 | 23 | |
| *, lower is better | 4 | 4 | 4 | |
| X, lower is better | 10 | 9 | 4 | |
| Precision p = O/(O + *) | 86% | 87% | 85% | |
| Recall r = O/(O + X) | 72% | 75% | 85% | |
| F-measure 2pr/( p + r) | 0.78 | 0.81 | 0.85 | |

is leaked. It works by performing differential black-box analysis on Android apps. MamaDroid [16], an Android malware detection system based on modeling the sequences of API calls as Markov chains. The system is designed to operate in one of two modes, with different granularities, by abstracting API calls to either families or packages. HornDroid [17], a tool for the static analysis of Android applications based on Horn clause resolution. HornDroid is the first static analysis tool for Android that comes with a formal proof of soundness covering a large fragment of the Android ecosystem.

### B. Dynamic Analysis

Dynamic analysis (Dynamic dataflow analysis) of an application involves testing and evaluation by feeding data to the application in real time within a controlled environment. It is used to analyze and monitor the situation behind the scenes while the application is running. Although static analysis tools are quick, they fail against encrypted, polymorphic and code transformed malware. Android app interaction is based upon UI elements and unsynchronized entry points; it is important to trigger to these events and elements to understand behavior. Android SDK is equipped with monkey runner, to automate gestures like swipe, key press, tap, etc.

TaintDroid [18] provides system-wide taint tracking for Android. It identifies data originating from sensitive sources like IMEI, camera, and GPS and monitors all network interfaces for sensitive data leak. Since Android L, TaintDroid can handle Android ART runtime. ARTDroid [19] could be extended to implement the dynamic taint analysis in the ART runtime in the future. ARTDroid is a hooking framework on Android ART runtime without modifications to both the Android system and the app's code. The effectiveness of dynamic dataflow depends on factors such as sink and source definition and input generation & test-driving. VetDroid [19] considers automation of source/sinks. It marks the information returned by functions calls backed by permissions. It means, it leverages the predefined Android permissions for automation. Machine learning techniques are used by SuSi to automatically identify sources/sinks in android APIs. Capper [20] instruments app instead of Android system for incorporating taint tracking. It has byte code rewriting approach to insert code in original one to keep track of data leakage. DroidScope [9] uses virtual machine introspection to mirror the three levels of an Android device: hardware, OS and Dalvik Virtual Machine facilitating the collection of detailed native and Dalvik instruction traces, profile API-level activity, etc. AASandbox executes the app in an isolated sandbox environment to analyze low-level interactions with the system. AntiMalDroid to detect Android malware that uses logged behavior sequence as the feature, and construct the models for further detecting malware and its variants effectively in runtime. CopperDroid [11] dynamically observe interactions between Android components and the underlying Linux system to reconstruct higher-level behavior.

## IV. RESULTS

DroidBench [12] is an Android-specific test suite. In version 2.0, 120 test cases are available and used to access both static and dynamic taint analysis problems, but in particularly has test cases for static analysis. Many have used DroidBench to measure and improve the effectiveness of their analysis tool. We compare the effectiveness of detecting data leakage of FlowDroid, AmanDroid and QARK tools and the results are in Table III. The results obtained are represented in terms of True Positive (O), False Positive (*), and False Negative (X). Multiple data leakages found can be represented in a single line, for example, Button2. QARK at present only supports LINUX environment but designers are working towards support for Windows. This tool can run in two modes, Scriptable mode and Interactive mode. Once an application apk is scanned for vulnerabilities the report is stored in html file in QARK directory (html file is re-written whenever tool is restarted or re-run for different apk). QARK represents it outcomes in form of potential vulnerability or warnings, illustrated as follows,

'*POTENTIAL VULNERABILITY - Access of phone number or IMEI, is detected in file: /home/…DroidBenchmaster/apk/Callbacks/Button3/classes_dex2jar/de/ecspride/Button1Listener.java.Avoid storing or transmitting this data*'.

This typical vulnerability was for the '*Button3.apk*' in the suite. If tools is able to detect a potential vulnerability it was represented as O, *, and X, and the number of warnings are simultaneously shown in Table III. The experiments were carried out in interactive mode for at least three times so that any sort of error in testing is eliminated.

## V. CONCLUSION

In this paper, we discussed the threats and security features along with tools for application security analysis. Android allows installation of third-party apks, which can be potentially harmful to the device, its user or user data. Attacks can range from data collection for targeted advertisement to potentially driven attacks for user-harm. Android requires additional security enforcements even if it has the presence of security features like user permissions, security at kernel level, and inter-process communication security. The two type of analyses, analyze source code as well as record the run-time behavior of application. We tested the performance of three static analysis tools using DroidBench including FlowDroid, AmanDroid and QARK. A comparison was presented which indicated that Amandroid has the highest precision rate (87%) while the highest recall vale of QARK (85%) resulted in overall highest f-measure value for QARK (0.85).